# Hot Carrier Dynamics in Photoexcited Gold Nanostructures: Role of Interband Excitations and Evidence for Ballistic Transport

Giulia Tagliabue[1,2], Adam S. Jermyn[3], Ravishankar Sundararaman[4], Alex J. Welch[1,2], Joseph S. DuChene[1,2], Artur R. Davoyan[1,5,6], Prineha Narang[7], Harry A. Atwater[1,2,6]

[1]Thomas J. Watson Laboratories of Applied Physics, California Institute of Technology, 1200 East California Boulevard, Pasadena, California 91125, USA
[2]Joint Center for Artificial Photosynthesis, California Institute of Technology, 1200 East California Boulevard, Pasadena, California 91125, USA
[3]Institute of Astronomy, Cambridge University, Cambridge CB3 0HA, UK
[4]Department of Materials Science and Engineering, Rensselaer Polytechnic Institute, 110 8th Street, Troy, New York 12180, USA
[5]Resnick Sustainability Institute, California Institute of Technology, Pasadena, California 91125, USA
[6]Kavli Nanoscience Institute, California Institute of Technology, Pasadena, California 91125, USA
[7]John A. Paulson School of Engineering and Applied Sciences, Harvard University, Cambridge, Massachusetts 02138, USA

Harnessing short-lived photoexcited electron-hole pairs in metal nanostructures has the potential to define a new phase of optoelectronics, enabling control of athermal mechanisms for light harvesting[1, 2], photodetection[3] and photocatalysis[4-6]. To date, however, the spatiotemporal dynamics and transport of these photoexcited carriers have been only qualitatively characterized. Plasmon excitation[7] has been widely viewed as an efficient mechanism for generating non-thermal 'hot' carriers[8-14]. Despite numerous experiments[3, 4, 15, 16], conclusive evidence elucidating and quantifying the full dynamics of hot carrier generation, transport, and injection has not been reported. Here, we combine experimental measurements with coupled first-principles electronic structure theory and Boltzmann transport calculations to provide unprecedented insight into the internal quantum efficiency, and hence internal physics, of hot carriers in photoexcited gold (Au) – gallium nitride (GaN) nanostructures. Our results indicate that photoexcited electrons generated in 20 nm-thick Au nanostructures impinge ballistically on the Au-GaN interface. This discovery suggests that the energy of hot carriers could be harnessed from metal nanostructures without substantial losses via thermalization. Measurements and calculations also reveal the important role of metal band structure in hot carrier generation at energies above the interband threshold of the plasmonic nanoantenna. Taken together, our results advance the understanding of excited carrier dynamics in realistically-scaled metallic nanostructures and lay the foundations for the design of new optoelectronic devices that operate in the ballistic regime.

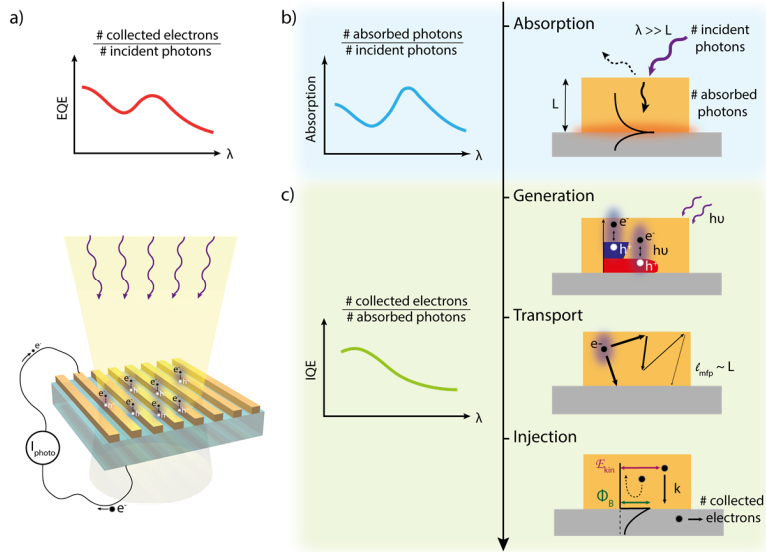

**Figure 1: Carrier generation and transport in photoexcited metal nanostructures.** a) Schematic representation of carrier generation and transport via internal photoemission (IPE) in a metal-semiconductor heterostructure (bottom): charge carriers created in the metal upon illumination are separated across the metal-semiconductor interface generating a photocurrent at sub-bandgap photon energies. Illustrative external quantum efficiency (EQE) spectrum (top), representing the wavelength (λ)-dependent photon-to-electron conversion probability. EQE can be decomposed into the product of absorption and internal quantum efficiency (IQE); b) Illustrative absorption spectrum (left) of a metal nanostructure displaying a resonant feature which can be engineered through photonic design. Plasmon excitation (right) yields high absorption in metallic nanostructures with characteristic dimension $L$ much smaller than the wavelength λ of the incident photon; c) Illustrative IQE spectrum (left) and schematic representation of the electronic processes which contribute to it (right), i.e. generation of carriers through intra- and interband transitions, propagation and scattering of the hot carriers with energy-dependent mean free path ($l_{mfp}$), and injection of hot carriers with adequate kinetic energy ($E_{kin}$) and momentum ($k$) across the Schottky barrier, $\Phi_B$.

Plasmon excitation enables tunable light absorption and photon confinement at nanoscale dimensions by coupling a photon with the collective oscillation of conduction electrons in the metal[1, 7]. Further, non-radiative plasmon decay forms a pair of excited 'hot' carriers[17] which can be injected from the metal into an adjacent semiconductor via internal photoemission[18-20] (IPE) to yield a photocurrent (Figure 1a). Using plasmonic nanostructures for light harvesting and charge carrier generation has thus enabled device miniaturization and introduced new functionalities in optoelectronic energy converters[3, 4, 15, 16, 21-24]. The reported device responsivity, or equivalently external quantum efficiency (EQE), has been shown to correlate with the plasmon resonance of the optical nanoantenna. While EQE describes the probability for photon-to-electron conversion (Figure 1a), more insight into the carrier dynamics can be obtained by determining the internal quantum efficiency (IQE), i.e., normalizing EQE by absorption. Whereas absorption measures the light-harvesting properties of the nanostructure (Figure 1b), IQE accounts for the electronic relaxation and transport processes within the device following photon absorption. For hot carriers, IQE is composed of three critical steps[25] (Figure 1c): (i) generation of a non-equilibrium distribution of 'hot' electrons and holes in the metal nanostructure upon plasmon decay via intraband (sp-sp) and interband (d-sp) optical transitions[17]; (ii) transport of these hot carriers to an interface either ballistically or via electron-electron and electron-phonon scattering and relaxation[17]; (iii) injection of carriers with appropriate momenta and sufficient kinetic energy above the interfacial Schottky barrier ($\Phi_B$)[25].

Although IQE is an established metric for semiconductor interband optoelectronic processes, to date, this fundamental characterization of device performance has received little attention in the context of hot carrier IPE systems. In fact, despite much theoretical progress in the understanding of individual electronic processes[17, 26-29], until very recently, an accurate description of the complex interplay of hot carrier generation and transport in realistic experimental structures, resolved by photon energy, has remained beyond the reach of computational tools[30]. As a result, the semi-classical Fowler theory continues to be used to interpret experimental IQE spectra[3, 22, 23, 31-35]. Failures of such approximation in the visible regime[31, 36, 37], however, have led to ad-hoc assumptions regarding the effects of plasmon excitation on specific electronic processes[32, 38]. Overall, limited model fidelity together with a lack of systematic experimental measurements have prevented a clear assessment of the physics underlying excited carrier transport and IPE. Our experimental studies of IPE hot electron collection across Au/n-GaN junctions as a function of incident photon energy, coupled with recently developed computational models[30] provide unprecedented insight of hot carrier generation, transport, and collection in metal-semiconductor Schottky junctions.

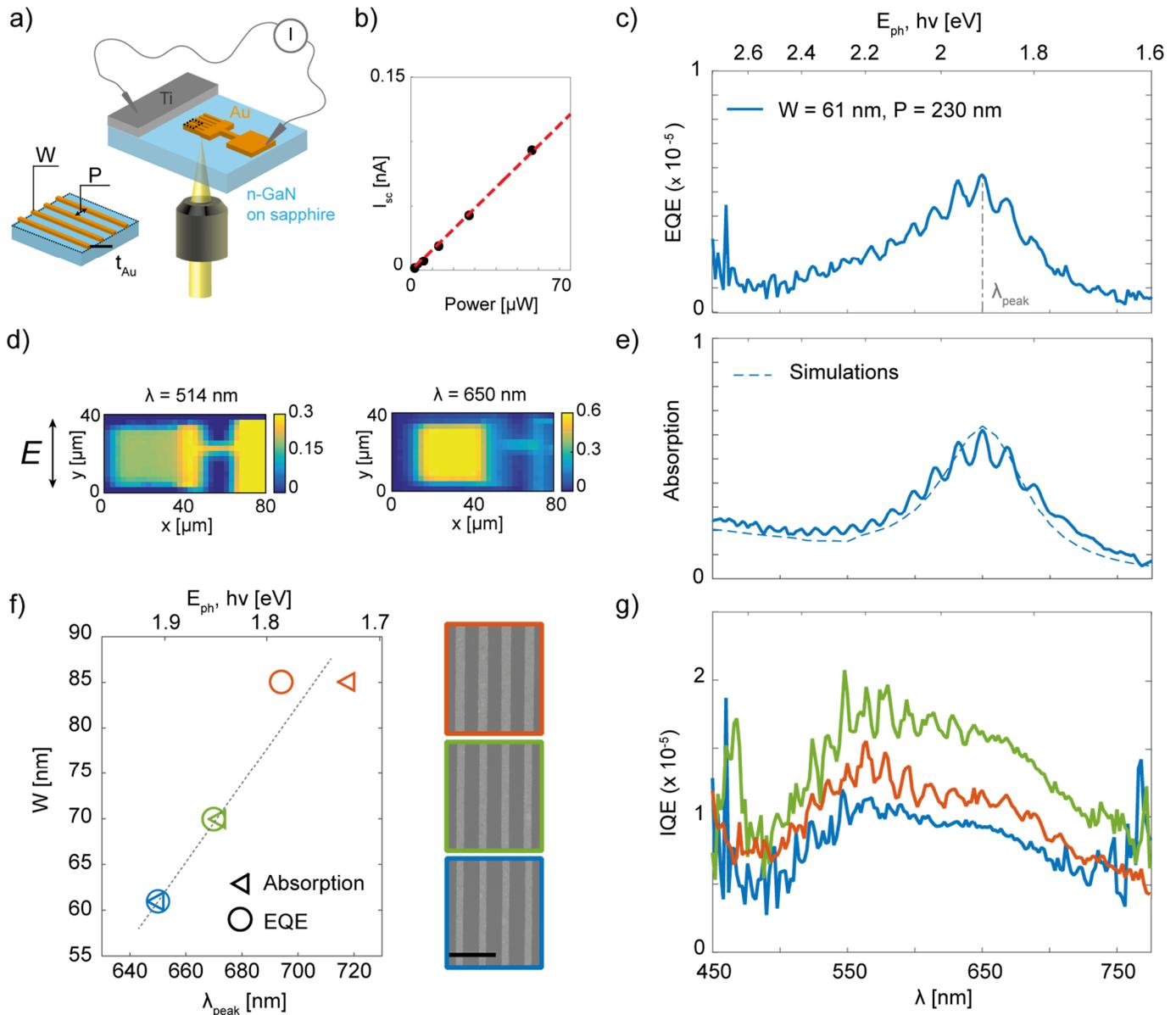

**Figure 2: Role of plasmon excitation on hot-electron IPE in metal-semiconductor heterostructures.** a) Schematic representation of the designed plasmonic heterostructures as well as measurement configuration: a 20 nm-thick, nano-patterned gold (Au) photoelectrode is fabricated on n-type GaN (3.4 eV band-gap, Schottky barrier $\Phi_B \sim 1.2$ eV) together with a 75 nm thick titanium (Ti) Ohmic contact; light is incident on the plasmonic resonant Au nanostripe array (stripe width $W$, array period $P$, see left inset) from the bottom and the photocurrent is collected via two microcontact probes; b) Short-circuit photocurrent $I_{sc}$ (i.e. 0 V applied bias) upon illumination of one heterostructure ($W = 61$ nm) with a diode laser ($\lambda_{laser} = 633$ nm) as a function of incident power; c) EQE spectrum of the fabricated heterostructure with stripe width $W = 61$ nm and periodicity $P = 230$ nm exhibiting a resonance peak at $\lambda_{peak} = 650$ nm; d) Spatial maps of absorption for illumination of the Au photoelectrode off-resonance (left, 514 nm – 2.14 eV) and on-resonance (right, 650 nm – 1.9 eV) with light polarized perpendicular to the stripes; e) Measured (solid line) and simulated (dashed line) absorption spectra for the same heterostructure exhibiting a plasmon resonance at $\lambda_{peak} = 650$ nm; f) EQE and absorption resonance peak wavelengths ($\lambda_{peak}$) for three heterostructures with constant array periodicity ($P = 230$ nm) and increasing nanostripe width, $W$, equal to 61 nm (blue), 70 nm (green) and 85 nm (red), respectively. Representative SEM micrographs are shown on the right (scalebar = 500 nm); g) IQE spectra of the three plasmonic heterostructures shown in part f).

To distinguish the role of plasmon excitation and material properties in plasmonic hot electron transport, it is necessary to compare experimentally-determined EQE, absorption, and IQE among heterostructures with distinct plasmon resonances but identical metal-semiconductor junctions. Accordingly, our experimental platform consists of planar Au plasmonic photodiodes on an optically-transparent yet electrically conductive n-type GaN substrate, allowing coupled electrical and optical (both transmission and reflection) characterization throughout the entire ultraviolet/visible/near infra-red (UV/VIS/NIR) spectral range. Each heterostructure consists of a large Au contact pad connected to an array of electrically conductive Au stripes, which serve as nanoantennas that support plasmon resonances in the VIS-NIR regime. For a fixed period ($P$) of 230 nm, specifically chosen to suppress any diffraction orders in the wavelength ($\lambda$) range of interest, the spectral position of the dipolar plasmon mode is controlled by adjusting the stripe width ($W$). Three hot-carrier heterostructures were constructed with $W$ of 61 nm, 70 nm and 85 nm to achieve plasmon resonances located at ca. 1.9 eV, 1.85 eV and 1.72 eV. The ultra-thin ($t_{Au}$ = 20 nm) dimension of the Au photoelectrode approaches the anticipated mean free path of hot carriers (ca. 10-20 nm around 2 eV[17]) and was chosen to maximize the ballistic collection of hot electrons without sacrificing optical absorption. A titanium (Ti) Ohmic contact completes the planar plasmonic diode so that photocurrent can be collected while illuminating the sample through the transparent substrate (Figure 2a and Methods).

The formation of a Schottky barrier ($\Phi_B$ ~ 1.2 eV[39], Figure 1d) at the Au/n-GaN interface ensures that electron-hole pair separation occurs even in the absence of an external bias. As expected, we observed a linear relationship between the short circuit photo-current, $I_{sc}$ and incident laser power (Figure 2b) when using a 633 nm diode laser to irradiate one stripe array ($W$ = 61 nm). We attribute the linear photoresponse to the injection of hot electrons from the Au nanoantennas into the n-GaN conduction band, since the incident photon energy is much less that the bandgap of the semiconductor ($E_g$ = 3.4 eV ~ 364 nm[40]),

For each heterostructure, EQE and absorption spectra are determined experimentally by measuring both the wavelength-dependent photocurrent as well as transmission and reflection spectra under the same illumination conditions of tunable, monochromatic light polarized perpendicular to the stripes (see Methods). For the heterostructure with $W$ = 61 nm, a resonance peak at $\lambda_{peak}$ = 650 nm can be observed in both spectra (Figure 2c,e), absorption being in excellent agreement with numerical simulations (Figure 2e, dashed line). Spatial maps of absorption in the photoelectrode were collected off-resonance above the interband threshold of Au ($\lambda$ = 514 nm < $\lambda_{IB}$ ~ 688 nm) as well as on-resonance ($\lambda_{peak}$ = 650 nm). In the first case, the unpatterned Au pad exhibits larger absorption than the array of nanoantennas (Figure 2b, left panel). Instead, on resonance (Figure 2b, right panel), absorption in the plasmonic stripe array ($\approx$ 60%) greatly exceeds that of the Au film. It is noted that this feature disappears upon rotating the incident light polarization by 90°. Such behavior confirms that the photocurrent originates from optical excitation of the dipolar plasmon mode in the nanoantennas. It is interesting to note that, not only the plasmon resonance, but also the fringes present in the absorption spectrum, which are due to Fabry-Perot interference[41] in the planar GaN/sapphire substrate structure, cause a modulation in the photocurrent response that is reproduced in the EQE spectrum.

Comparing the optical (absorption) and electrical (EQE) performance of three hot carrier heterostructures with varying stripe width, we find a close correlation between the plasmon excitation wavelength and the EQE peak response (Figure 2f). Increasing $W$ from 61 nm to 85 nm red-shifts both absorption and EQE peak positions ($\lambda_{peak}$) by a commensurate amount. In contrast, the IQE spectra, determined by taking the ratio of EQE and absorption (Figure 2g), do not exhibit any spectral feature that can be directly associated with the characteristic plasmon peak wavelength of each device. The striking similarity of the three IQE curves shows that the role of plasmon excitation in enhancing the photodiode performance (i.e. EQE) is primarily photonic in nature. That is, tunable plasmon resonances efficiently couple far-field radiation into nanoscale volumes and this mechanism dominates the EQE across a broad range of wavelengths. Yet this increased light-harvesting capability does not alter carrier generation rates or the internal device physics governing carrier transport. This observation implies that the intrinsic material properties of the metal are regulating the electronic response of the heterostructure and that plasmon excitation does not *a priori* selectively enhance the rate of any particular decay process or transport mechanism. Interestingly, as also remarked in previous studies[31, 32, 36], we observed that all three IQE curves were characterized by a broad, asymmetric feature peaking around 560-565 nm (~2.2 eV), which cannot be described by conventional Fowler models for IPE. Contrary to previous speculations about the role of indirect bandgap materials[36], our results on a direct bandgap semiconductor (n-GaN) indicate that it is the electronic band structure of the metal that determines the energy dependence of the IQE.

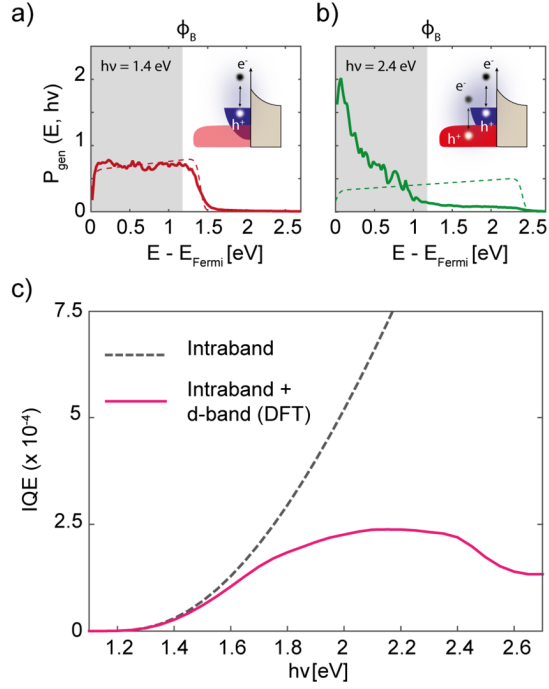

**Figure 3: Impact of inter- and intraband transitions on IQE.** a) Prompt hot electron energy distribution ($P_{gen}$) showing the carrier energy $E$ above the Au Fermi level ($E_F$) calculated with DFT (solid line) as well as under the parabolic-band approximation (Fowler-like model, dashed line) for incident photon energies ($h\nu$) of 1.4 eV and b) 2.4 eV. The shaded area in both plots depicts the position of the Schottky barrier, $\Phi_B$, limiting the possibility of collection to those carriers with energy $E-E_F > \Phi_B$. The insets show a schematic of the metal and semiconductor band structure illustrating the predominance of intraband transitions (a) and co-existence with interband transitions (b) as well as the presence of the Schottky barrier at the interface; c) IQE spectra calculated based on the $P_{gen}$ obtained with DFT (purple solid curve), i.e. including interband transitions, as well as with parabolic-band approximation (grey dashed curve), i.e. accounting only for intraband transitions. For the injection process, conservation of tangential momentum is assumed[22]. Transport of hot electrons within the metal nanostructure has been neglected.

We relate the particular shape of the IQE curves to the interplay between the two hot carrier generation mechanisms, namely intraband and interband transitions, as well as their corresponding hot carrier distributions relative to the Schottky barrier height present at the metal-semiconductor interface. The interband and intraband decay rates are determined from density functional theory (DFT) calculations, which generate the prompt hot-electron-energy distribution. The decay rate is strongly dependent on both incident photon energy and the electronic band structure of the metal[17, 27]. For photon energies below the interband threshold of Au ($h\nu_{IB} \sim 1.8$ eV) hot electrons generated via intraband transitions have a nearly uniform probability at all energies from the Fermi level up to the photon energy (Figure 3a, solid line). As a result, intraband excitation creates a sizable fraction of the hot electron distribution at energies above the Schottky barrier height (grey shaded area in Figure 3a). In this low photon energy regime there is very good agreement between the Fowler model, based on the parabolic-band approximation, and full DFT calculations (compare solid line with dashed line in Figure 3a). On the other hand, above $h\nu_{IB}$ (Figure 3b, solid line), a much higher probability distribution of low energy carriers is observed, since hot electrons originate from d-band levels deep below the Fermi level. Consequently, there is a substantial reduction in the fraction of high-energy electrons created from intraband transitions compared to that predicted for the case of a purely parabolic band (dashed line). This interplay, combined with the height of our Schottky barrier (grey shaded area in Figure 3a,b), results in a reduction in IQE at energies above the interband threshold (Figure 3c, solid line). This is in sharp contrast to predictions of the Fowler model, which accounts exclusively for intraband processes (Figure 3c, dashed line). However, it must be recognized that even above $h\nu_{IB}$ both types of transitions occur simultaneously and high-energy carriers continue to be generated, though with decreasing probability. Indeed, plasmon excitation does not alter this interplay, as it does not directly influence the hot carrier distribution, only the number of absorbed photons at a given frequency. The change in the dominating optical transition mechanism with increasing photon energy explains why the metal band structure, and in particular its interband threshold, has such a profound effect on the overall IQE of hot-carrier devices.

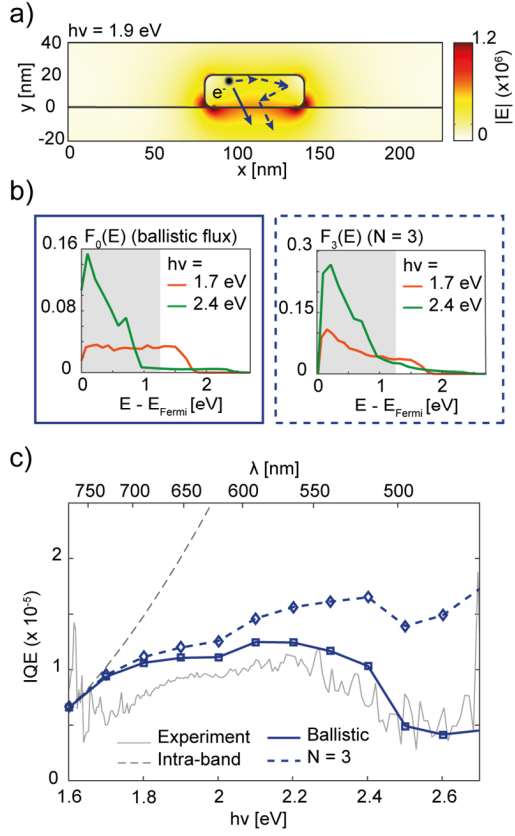

**Figure 4: Hot electron generation and transport in plasmonic nanoantennas.** a) Calculated spatial profile of the electric field norm |E| at resonance ($\lambda = 650$ nm, $h\nu = 1.9$ eV) for the experimental structure with $W = 61$ nm and $P = 230$ nm. |E| in the metal defines the spatial generation profile of the hot carriers. As schematically illustrated, hot electrons then propagate across the metal structure and reach the Au-GaN interface either ballistically (solid arrow) or after scattering (dashed arrows); b) Energy-resolved flux of hot electrons reaching the Au-GaN interface for photon energy of 1.7 eV (orange curves, weak interband contribution) and 2.4 eV (green curves, strong interband contribution). The shaded area shows the position of the Schottky barrier. The left plot shows the flux of carriers reaching the interface ballistically while the right plot is the flux of carriers including those which have undergone up to $N = 3$ scattering events; c) IQE spectra calculated based on the computed energy-resolved fluxes, both for the ballistic case (blue solid curve) and for $N = 3$ (blue dashed curve), under the assumption of tangential momentum conservation for the injection probability[22]. The grey dashed curve represents the IQE estimated based on the fit of Fowler yield, $IQE_{Fowler} = C \cdot (h\nu - \Phi_B)^2 / h\nu$ with $\Phi_B \sim 1.2$ eV and $C = 6.7 \times 10^{-5}$. The grey solid curve is the experimentally determined IQE (Figure 2g, blue curve).

A microscopic understanding of hot carrier transport in Au/n-GaN heterostructures is obtained by comparing experimental measurements to results of a recently-developed theoretical framework that combines electromagnetic simulations, *ab initio* DFT calculations, and Boltzmann transport methods to compute the generation and transport of hot carriers within realistically-scaled (ca. 10-100 nm) metallic structures[30] (see Methods). From electromagnetic simulations, we first determine the electric field profile in a single Au nanoantenna ($W = 61$ nm, Figure 4a). The initial energy and momentum distribution of the hot carriers are obtained from plasmon decay rates and electronic optical excitations derived from DFT calculations[17, 27], which account for the anisotropies associated with these quantities in the interband regime. Energy-dependent lifetimes and mean free paths ($l_{mfp}$) are also calculated with *ab-initio* methods accounting for both electron-electron and electron-phonon scattering processes[26, 42]. This information is combined in a Boltzmann transport calculation[30] where we compute the propagation of carriers across the Au nanostructure, determining changes to their energy distribution as well as the number of scattering events they experience. For each photon energy, out calculations yield the energy-resolved flux $F_N(E)$ of hot electrons with energy $E$ above the metal Fermi level that reach the Au/n-GaN interface after up to $N$ scattering events. Attesting to the validity of our computational approach, the energy-resolved flux of hot electrons that reach the interface ballistically, $F_0(E)$ (Figure 4b, left panel), retains the key features described in Figure 3a, left panel. The model also shows that scattering processes serve to homogenize the hot carrier distributions by smoothing the transition between the intra- and interband generated carriers that reach the interface (Figure 4b, right panel).

Estimating the injection probability, $P_{inj}(E)$ across the Schottky barrier based on the assumption of tangential momentum conservation[22], we then calculate IQE as:

$$IQE = \int_{\Phi_B}^{\infty} F_N(E) \cdot P_{\text{inj}}(E) dE \ \ for \ N = 0, 1, ..$$

The blue solid curve in Figure 4c represents the IQE spectrum obtained from $F_0(E)$ and the blue dashed curve is the one obtained from $F_3(E)$. Including additional scattering events only changes the IQE by 0.01%, indicating that the vast majority of hot electrons undergo no more than three scattering events before being collected. Overall, our parameter-free model of carrier generation, transport, and injection is in excellent quantitative agreement with the experimental data (grey solid curve). This is the first time that a detailed description of material properties and device geometry for metal nanostructures was able to capture the details of plasmonic hot carrier transport under illumination, both on and off resonance. Strikingly, the results of our model indicate that more than 90% of the hot carriers are collected ballistically at photon energies below 2 eV ($\lambda > 620$ nm), implying that hot carrier transport in our Au nanoantennas occurs in the ballistic regime at the plasmon peak position. Strong electric field confinement near the metal-semiconductor interface indicates that hot carriers are preferentially generated near the site of collection (Figure 4a) and an expected $l_{\text{mfp}}$ of ca. 10-20 nm for hot electrons ca. 2 eV above the metal Fermi level[17] further implies that such ballistic transport is feasible in our antenna structures. Therefore, by enabling strong light localization in metallic nanostructures with dimensions comparable with the short, energy-dependent $l_{\text{mfp}}$ of hot electrons in metals, plasmon excitation may be able to realize optoelectronic systems that operate in the ballistic regime.

To summarize, our combined experimental and theoretical studies of IQE in plasmonic nanoantennas reveal that plasmon excitation enables the efficient coupling of far-field radiation into nanoscale volumes, but does not alter the fundamental physics governing the operation of hot-carrier devices. Instead, detailed analysis of the IQE spectra emphasize the role of interband and intraband decay processes, as well as nanometer-scale transport, in determining the distribution of hot carriers that are actually collected via IPE. Significantly, our observation of ballistic electrons is encouraging for efforts to use ballistic hot-carrier collection for ultra-fast photodetection and excited-state photocatalysis. More generally, these results indicate the practicality of elucidating and tailoring the IQE of hot carrier devices based on precise knowledge of the electronic processes involved in the IPE mechanism. Further experiments using time-resolved measurements may expand our understanding of hot carrier transport, and allow for more comprehensive comparison with theoretical predictions. As an outlook, the agreement between our experimental data and detailed theoretical hot carrier transport model suggests that this combined approach can be a powerful tool to guide the design of future hot carrier optoelectronic devices.

## Methods

*Sample fabrication*

GaN films on sapphire were purchased from Xiamen (4 ± 1 μm thick GaN layer, Ga-face, epi-ready, $N_d$ = 5-7*10$^{17}$ cm$^{-3}$, $\rho$ < 0.5 Ω·cm, < 10$^8$ dislocations/cm$^2$). A layer of S1813 was spin coated on the substrate (40 s, 3000 rpm) and post-baked for 2 min at 115 °C. The Ohmic pattern was exposed for 40 s and then developed for 10 s in MF319®. Then, 75 nm of Ti were deposited with e-beam evaporation (1.5 Å/s, base pressure lower than 5 x 10$^{-7}$ Torr). A layer of PMMA 495-A4 was spin coated on the sample (1 min, 4000 rpm) and baked for 2 min at 180 °C. Next, a layer of PMMA 950-A2 was spin coated on top of it (1 min, 5000 rpm) and also baked for 2 min at 180 °C. Then, e-beam lithography was used to write the nanoantenna pattern (Quanta FEI, NPGS system). Beam currents of approximately 40 pA were used with exposures ranging from 350 μC/cm$^2$ to 500 μC/cm$^2$, thus achieving different stripe widths with equal pitch. A 20 nm Au layer was then deposited with e-beam evaporation (Lesker) (0.8 Å/s, base pressure lower than 2 x 10$^{-7}$ Torr). Importantly, before any metal deposition, the sample was exposed to a mild oxygen plasma (30 s, 200 W, 300 mT) to remove any photoresist residual, dipped in a 1:15 NH$_4$OH:DI H$_2$O solution for 30 s to remove any surface oxide layer and finally rinsed in water (30 s) and blown dry with N$_2$ gas. The substrate was then immediately loaded into the e-beam evaporator chamber, minimizing the time of exposure to ambient atmosphere.

*Photocurrent Measurements*

A Fianium laser (2 W) was used as the light source for plasmon excitation. The beam was monochromated (slit width 200 μm), collimated, and finally focused onto the sample with a long working distance, low-NA objective (Mitutoyo 5X, NA = 0.14). A Si photodetector was used to measure the transmitted power or, using a beam splitter, the reflected power incident on the sample. A silver mirror (M, Thorlabs) was used to normalize the reflection and the background (BG) was subtracted from all the measurements. A tilted glass slide was used to deflect a small amount of incident power from the laser onto a reference photodiode for coincident recording of the laser power incident on the sample. A chopper, typically at a frequency of ~100 Hz, was used to modulate the incident power and thus the photocurrent signal, which was subsequently processed with a lock-in amplifier. An external, low-noise current-to-voltage amplifier was used to feed the signal to the lock-in. Piezoelectric micro-probes (Mibots®) are utilized to electrically contact the sample and perform all of the photocurrent measurements.

*Numerical Simulations*

A commercial finite element method software (COMSOL) is used to perform the electromagnetic simulations. The 3D simulations are performed to estimate absolute absorption values as well as 3D internal electric field distributions to be used in the subsequent hot carrier generation and transport code. The scattered field formulation is utilized. For the background field calculation, a port boundary condition with excitation "ON" is used to launch a plane wave with normal incidence and variable wavelength as well as for the recording of the reflected wave. A second port boundary condition without excitation is used to record the transmitted wave. Perfect magnetic conductor and periodic boundary conditions are used on the side-walls (width of the unit cell equal to the array pitch, $P$ = 230nm, length of the cell equal to 50 nm). For the calculation of the scattered field, perfect-matched layers (PMLs) are used in place of the port boundary conditions.

*Hot carrier generation and transport predictions*

The hot carrier flux is computed by iteratively evaluating the effects of transport and scattering. In each iteration, transport effects are computed using the 1D Green's function (exp(-$x/l_{mfp}$), where $l_{mfp}$ is the mean-free path) on a tetrahedral mesh. Multiple different directions are integrated via Monte Carlo sampling. This results in a deposition of transported carriers at the surface and scattered carriers in the interior. The scattered carriers are then transformed via the scattering matrix elements to produce a new energy distribution at each point in the mesh, which is used as the input to the next round of transport calculations. The initial input distribution is obtained using the carrier-energy resolved dielectric function Imε($\omega,E$) and the input electromagnetic field from COMSOL, evaluated on the same tetrahedral mesh. Imε($\omega,E$) and the energy-dependent mean-free path $l_{mfp}(E)$ are obtained using Fermi's golden rule, with electron-phonon and electron-photon matrix elements calculated using the DFT software JDFTx. See Ref 30 for further details.

**Acknowledgments**

This material is based upon work performed by the Joint Center for Artificial Photosynthesis, a DOE Energy Innovation Hub, supported through the Office of Science of the U.S. Department of Energy under Award No. DE-SC0004993. R.S., A.S.J., and P.N. acknowledge support from NG NEXT at Northrop Grumman Corporation. Calculations in this work used the National Energy Research Scientific Computing Center, a DOE Office of Science User Facility supported by the Office of Science of the U.S. Department of Energy under Contract No. DE-AC02-05CH11231. A.D. and H.A.A. acknowledge support from the Air Force Office of Scientific Research under grant FA9550-16-1-0019. G.T. acknowledges support from the Swiss National Science Foundation through the Early Postdoc Mobility Fellowship, grant n. P2EZP2_159101. P.N. acknowledges support from the Harvard University Center for the Environment (HUCE). A.S.J. thanks the UK Marshall Commission and the US Goldwater Scholarship for financial support. AJW acknowledges support from the National Science Foundation (NSF) under Award No. 2016217021.